Automatic detection of key innovations, rate shifts, and diversity-dependence on

phylogenetic trees

Daniel L. Rabosky

Department of Ecology and Evolutionary Biology and Museum of Zoology, University of

Michigan, Ann Arbor, Michigan  48130 USA

Email: drabosky@umich.edu

Short title: Modeling macroevolutionary mixtures on phylogenies

Abstract

15    A number of methods have been developed to infer differential rates of species diversification through time and among clades using time-calibrated phylogenetic trees.

However, we lack a general framework that can delineate and quantify heterogeneous

mixtures of dynamic processes within single phylogenies. I developed a method that can

identify arbitrary numbers of time-varying diversification processes on phylogenies

20    without specifying their locations in advance. The method uses reversible-jump Markov

Chain Monte Carlo to move between model subspaces that vary in the number of distinct

diversification regimes. The model assumes that changes in evolutionary regimes occur

across the branches of phylogenetic trees under a compound Poisson process and

explicitly accounts for rate variation through time and among lineages. Using simulated

25    datasets, I demonstrate that the method can be used to quantify complex mixtures of

time-dependent, diversity-dependent, and constant-rate diversification processes. I

compared the performance of the method to the MEDUSA model of rate variation among

lineages. As an empirical example, I analyzed the history of speciation and extinction

during the radiation of modern whales. The method described here will greatly facilitate

30    the exploration of macroevolutionary dynamics across large phylogenetic trees, which may have been shaped by heterogeneous mixtures of distinct evolutionary processes.

Introduction

35    Perhaps the most general feature of biological diversity on Earth is the extent to which it

varies - either through space, through time, or among different kinds of organisms.

Biologists have long been fascinated by the observation that some groups of organisms

contain far more species than other groups. For example, within vertebrates, lineages

such as tetrapods (22000+ species), therian mammals (5000+ species), and teleosts

40    (30000+ species) are several orders of magnitude more diverse than their respective sister clades (lungfishes, 6 species; monotremes, 5 species; holosteian fishes, < 10 species).

This phylogenetic variation in species richness is mirrored by analogous variation in diversity through time. Paleontological evidence indicates that species richness has undergone dramatic changes during the past 550 million years [1,2]. Finally,

45    contemporary species richness varies dramatically among geographic and climatic

regions [3,4]. At least in part, the causes of phylogenetic, temporal, and spatial variation

in species richness are thought to reside in the evolutionary processes of speciation and

extinction. Consequently, there has been great interest in studying historical patterns of

species diversification through time, towards understanding how and why speciation and

50    extinction rates might vary through time, through space, and among clades [5,6,7,8,9].

The fossil record has provided insight into the temporal dynamics of species diversification [10,11], but analyses have generally been restricted to groups with exceptional fossil records and/or to relatively coarse temporal and phylogenetic scales.

55    Because of the difficulties in applying paleontological approaches to many groups of organisms that lack adequate fossil records, there is great interest in extracting information about macroevolutionary dynamics from time-calibrated phylogenetic trees of extant species only [8,12]. The increase in the availability of such phylogenies has helped catalyze a surge of methodological [13,14,15] and meta-analyses [16,17,18,19] on

60  the temporal dynamics of speciation and extinction through time. At the same time, a range of new approaches have been developed to assess the extent to which rates of species diversification vary among lineages [20,21] or in association with character states [22,23,24,25].

65  To date, few macroevolutionary studies have simultaneously accounted for rate variation through time and among lineages [6,13,26]. Increasing evidence suggests that failing to accommodate rate variation through time and among lineages can lead to profoundly biased parameter estimation [27] and conceptually flawed interpretations of the factors that regulate species richness within clades or regions [26,28].

70

In this article, I introduce a new framework for studying patterns of rate variation through time and among lineages using time-calibrated phylogenies of extant species. The approach is premised on the idea that phylogenetic trees are frequently shaped by heterogeneous mixtures of distinct processes. For example, some phylogenies may reflect

75  mixtures of both diversity-dependent and constant-rate diversification processes (Figure 1). There is already considerable evidence that many empirical phylogenies have been shaped by multiple distinct evolutionary processes [13,29], and challenges of modeling such data are expected to increase with phylogenetic tree size.

80    My general approach assumes that shifts between macroevolutionary regimes occur across the branches of a phylogenetic tree under a compound Poisson process. This framework has been used previously to model among-lineage variation in rates of molecular evolution [30]. The number of such transitions between distinct processes is assumed to follow a Poisson distribution. Rather than assume a fixed number of distinct

85    processes on a given phylogenetic tree, I use reversible jump Markov Chain Monte Carlo [31] (hereafter, rjMCMC) to automatically explore the universe of models that differ in the number of distinct evolutionary regimes. The method thus enables exploration of a vast state space of possible models to explain a given phylogenetic diversification pattern.

90    The method described here differs from previous methods in several key respects. First, the method does not assume that rates of speciation and extinction are constant through time within clades, thus relaxing the assumption of time-homogeneous diversification used in most previous multi-model approaches [20,32]. Second, the location and number of distinct evolutionary processes ("regimes") represent random quantities that are

95    themselves estimated from the data. In addition, by adopting a Bayesian approach, we can algorithmically explore a greater number of candidate models than is possible with incremental (e.g., stepwise) information-theoretic approaches [20]. Because rjMCMC samples diversification models in proportion to their posterior probability [31,33], model selection emerges automatically from the analysis. Finally, the method provides marginal

100    distributions of speciation and extinction rates for every branch in a phylogenetic tree.

Materials and Methods

105    Compound Poisson process model of diversification rate variation

The model assumes that phylogenetic trees are shaped by a countable set of distinct and

potentially dynamic evolutionary processes of speciation and extinction. Transitions

between processes, or "events", are assumed to occur across the branches of the

110    phylogeny under a compound Poisson process [30]. Let $\xi_i$ denote the mapping of the i'th

transition to a specific location on the tree; thus, $\xi$ denotes a unique location on a specific

branch of the tree. Nodes and branches descended from a mapped transition $\xi_i$ inherit the

evolutionary process, denoted by $\Phi_i$, that begins at point $\xi_i$. The process $\Phi_i$ terminates at

terminal branches, or at the next downstream transition (Figure 1). Thus, the occurrence

115    of a transition defines a connected subgraph of adjacent nodes, but does not necessarily include all of the descendent nodes downstream of a particular transition.

Any tree is necessarily governed by at least one process that begins at the root node and

the number of additional transitions is a Poisson-distributed random variable with rate

120    parameter $\Lambda$. In the MCMC implementation of the model described below, new

transitions can be added to the tree, and existing transitions can be moved or deleted from

the tree. The addition of a transition results in a new evolutionary process that is

decoupled from the parent process. For example, consider a phylogeny with dynamics

governed by just a single process, $\Phi_R$. If a transition occurs at position $\xi_i$, then all lineages

125    downstream from point ξ$_i$ are governed by a new evolutionary process, $\Phi_i$. Formally,

each possible count of transitions defines a diversification model, and we denote a model with $k$ distinct transitions by M$_k$. The addition of a process to a tree with $k$ transitions thus entails a jump from model M$_k$ to M$_{k+1}$. There is no upper bound on the number of transitions, as multiple transitions can occur on a given branch. The minimum model,

130    with a single process, corresponds to $k = 0$ and contains zero transitions.

I assume that each process $\Phi$ represents a distinct time-varying process of speciation ($\lambda$)

and a constant background rate of extinction ($\mu$). We used an exponential change function

to model variation in speciation rates through time within a particular process, such that

135

$$\lambda\left(t_i\right) = \lambda_{0,i} \exp\left(z_i t_i\right) \tag{1}$$

where $\lambda$ is the rate of speciation for a process at time t$_i$ relative to the start of the process

and where $\lambda_{0,i}$ and $z_i$ represent the initial speciation rate and rate change parameter for the

140    i'th transition. For notational clarity, parameters associated with the root process are

denoted with zero subscripts: for example, $\lambda_{0,0}$ and $z_0$ correspond to the initial speciation

rate and rate-change parameter for the process associated with the root of the tree. This

model is equivalent to the SPVAR model from Rabosky and Lovette [14]. The

exponential change function is a natural choice for modeling both time-varying and

145    diversity-dependent speciation, because an exponential change in speciation with respect to time closely approximates a linear change in speciation with respect to diversity [34].

The full model thus includes the possibility that a single time-varying diversification process describes the entire phylogeny [14] as well as the possibility that many independent time-varying processes govern evolutionary dynamics across the tree.

150

I implemented the model in a Bayesian framework. Bayesian approaches have already

been used effectively to model single processes on phylogenetic trees [35,36], but in this

case, the number of distinct processes is itself a random quantity. I constructed a

transdimensional Markov chain that could move between models containing different

155    numbers of processes. This is known as "reversible jump" Markov Chain Monte Carlo

(rjMCMC), as it involves probabilistic "jumps" between model subspaces of different

dimensionality [31]. An attractive feature of this approach is that the Markov chain

samples diversification models in proportion to their posterior probability. Thus, the

relative probabilities of diversification models with (0, 1, 2, 3.... $k$) distinct processes can

160    be computed immediately by tabulating the relative frequencies of those models in the

MCMC output. Several recent studies have used rjMCMC to study variation in rates of phenotypic evolution across phylogenetic trees [37,38].

Bayesian implementation

165

The full model contains parameters for the overall rate at which transitions occur ($\Lambda$), as

well as location ($\xi$) and diversification parameters ($\lambda, z, \mu$) for each transition. I simulated

a Markov process that (i) permitted incremental transitions to new diversification models

($M_k \rightarrow M_{k+1}$ or $M_k \rightarrow M_{k-1}$), and (ii) updates to parameters of the current models.  Note

170    that my usage of the word model in this context can refer to either the overall compound

Poisson process model, or to submodels with distinct numbers of processes (e.g., $M_1$, $M_2$, ... $M_N$). The Markov chain is updated using the following moves: (1) a transition is added to the tree, (2) a transition is deleted from the tree, (3) the position of an existing transition ($\xi_i$) is updated, (4) the rate at which transitions occur is updated ($\Lambda$), (5) the

175   initial speciation rate for the i'th transition is changed ($\lambda_{0, i}$), (6) the rate-change parameter for the i'th transition ($z_i$) is updated, and (7) the extinction rate for the i'th transition is changed ($\mu_i$).

For within-model moves that do not involve changes in the dimensionality of the full

180   model, acceptance probabilities follow the standard Metropolis-Hastings formulation [39,40], or

$$\min\left\{1, \frac{f(\theta')}{f(\theta)} \frac{\pi(\theta')}{\pi(\theta)} \frac{q(\theta\mid\theta')}{q(\theta'\mid\theta)}\right\}$$

(2)

where $\theta$ and $\theta'$ are parameter vectors corresponding to current and proposed states, $f(\cdot)$ and $\pi(\cdot)$ are the corresponding likelihood and prior densities, and $q(\theta' \mid \theta)$ is the

185   relative probability of proposing a move to parameter vector $\theta'$ given that the current state is $\theta$.

The acceptance probability for moves that transition between models requires a more general formulation [31,41], of which standard Metropolis-Hastings is a special case. In

190   the present framework, we propose to jump from some model $M_k$ with parameter vector $\theta$ to a new model $M_{k+1}$ with parameter vectors $\theta'$ and $\psi$, where $\theta$ denotes parameters that are common to both models and $\psi$ denotes parameters that occur in the proposed model but not the current model. To move between models, we generate a random vector $\nu$ from some known density, $q(v)$. We then map the current state and the random vector to the

195   new state ($\theta'$, $\psi$) through use of a mapping function g($\theta$, $v$). The random vector $v$ has a

number of elements equal to the number of parameters in $\psi$, thus satisfying the

dimension-matching requirement for transdimensional moves. The acceptance probability

for this move is given by

$$\min\left\{1, \frac{f(\theta',\psi)}{f(\theta)} \frac{\pi(\theta',\psi)}{\pi(\theta)} \frac{\pi(M_{k+1})}{\pi(M_k)} \frac{q(M_k \mid M_{k+1})q(\theta \mid \theta')}{q(M_{k+1} \mid M_k)q(\theta' \mid \theta)q(v)} \left| \frac{\partial g(\theta,v)}{\partial(\theta,v)} \right| \right\}$$   (3)

200   where q($M_k \mid M_{k+1}$) denotes the probability of proposing a move from model $M_{k+1}$ to model $M_k$, $\pi(M_k)$ is the prior probability of model $M_k$, and the last term is the determinant of the Jacobian matrix for the transition from the vector ($\theta$, $v$) to ($\theta'$, $\psi$) via the mapping

function $g(\,\cdot\,)$. The corresponding reverse move is deterministic and the acceptance

probability is given by the inverse of the numerator in equation 3, with the exception of

205   the case where $k$ equals 0 or 1 (discussed below).

In the model described here, an increase from $M_k$ to $M_{k+1}$ involves the addition of four new parameters to the process: $\psi = (\xi_{k+1}, \lambda_{0,k+1}, z_{k+1}, \mu_{k+1})$. During model-jumping proposals, all parameters $\theta$ are mapped to $\theta'$ via the identity function, such that $\theta' = \theta$.

210   The mapping from $v$ to $\psi$, or g($v$), was also defined using simple identity relationships:

$$\begin{aligned} \xi_{k+1} &= v_1 \\ \lambda_{0,k+1} &= v_2 \\ z_{k+1} &= v_3 \\ \mu_{k+1} &= v_4 \end{aligned} \quad .$$   (4)

Variables $v_1$, $v_2$, $v_2$, and $v_3$ were sampled from the corresponding prior distributions for $\xi$,

$\lambda_0$, $z$, and $\mu$. The Jacobian term reduces to the identity matrix and has a determinant of 1.

215 Under the compound Poisson process, the overall (whole-tree) rate at which transitions occur under the model is $\Lambda$. The prior ratio for models $M_{k+1}$ to $M_k$, given $\Lambda$, is simply the ratio of Poisson densities with $k+1$ and $k$ transitions, or

$$\frac{\pi\left(M_{k+1} \mid \Lambda\right)}{\pi\left(M_k \mid \Lambda\right)} = \frac{\left(\Lambda\right)^{k+2} e^{-\Lambda} \big/ (k+1)!}{\left(\Lambda\right)^{k+1} e^{-\Lambda} \big/ k!} = \frac{\Lambda}{k+1} \quad . \qquad (5)$$

220 To compute the proposal probability $q(M_k \mid M_{k+1})$, let $d_{k+1}$ represent the probability of making a move that deletes one of $k + 1$ transitions on the tree, and let $b_k$ represent the probability of adding a transition when there are currently $k$ transitions on the tree. The probabilities $d_k$ and $b_k$ are typically equal to 0.5, since addition and deletion moves are equiprobable for most values of $k$. However, when the tree includes just a single process

225 and no transitions ($k = 0$), the relative probability of adding a transition is equal to 1.0, as the root process cannot be deleted. In this case, $d_1 = 0$ and $b_1 = 1$, because only additions of transitions (and not deletions) can be proposed. This leads to an asymmetrical proposal ratios for adding transitions (when $k = 0$) and for deleting transitions (when $k = 1$). When $k = 0$, the ratio $d_2 / b_1$ is equal to 0.5. When there are two processes on the tree ($k = 1$), the

230 proposal ratio is asymmetrical and $b_{k-1} / d_k$ is equal to 2.0. This compensates for the excess of gain proposals that occur when there is just a single process on the tree. Otherwise, $q(M_{k+1} \mid M_k) = q(M_k \mid M_{k+1}) = 1$.

Because elements of $v$ are sampled from prior distributions for $\xi$, $\lambda_0$, $z$, and $\mu$, the prior on

235 $\psi$ in the numerator of equation 3 is equal to the density $q(v)$ in the denominator, as in [30].

The acceptance probability for the addition of a transition is thus a function of the likelihood ratio, the prior ratio for models $M_{k+1}$ and $M_k$, and the proposal ratio for the models, or

$$\min\left(1, L\frac{\Lambda T d_{k+1}}{(k+1)b_k}\right) \qquad\qquad (6)$$

240

where $L$ is the likelihood ratio of current and proposed states.

The acceptance probability for a move that deletes one of $k$ transitions from the tree involves inverting the ratio term from equation (6) and modifying subscripts to reflect the

245 fact that we are proposing a move to a state with $k$ - 1 transitions. The proposal ratio becomes $b_{k-1} / d_k$, leading to an overall acceptance probability of

$$\min\left(1, L\frac{k b_{k-1}}{\Lambda T d_k}\right). \qquad\qquad (7)$$

250 The positions of transitions were updated using global and local moves. A global move entailed sampling a new map location $\xi$ from tree and allowed transitions to shift to any point on the tree with uniform probability. A local move involved shifting the position of a transition by a small random quantity that was sampled from a uniform distribution. I fixed the ratio of global:local proposals at 1:10 for all analyses described here. The

255 acceptance probability for a move that changes the position of a transition is equal to $\min(1, L)$.

To update any of the zero-bounded rate parameters in the model ($\Lambda$, $\lambda_i$, $\mu_i$), I used a proportional shrinking-expanding proposal [42], such that

260

$$r' = re^{\eta(U-0.5)} \qquad (8)$$

where $r$ is the current value of the rate parameter, $U$ is a random variable sampled from a uniform $(0, 1)$ distribution, and $\eta$ is a tuning parameter. The acceptance probability of a

265 move that updates such a rate parameter is

$$\min\left(1, Le^{\eta(U-0.5)}\right) \qquad (9)$$

Finally, I used a sliding window proposal to update the value of the rate change

270 parameter $z_i$. Here, a random variable is sampled from a uniform $(-\delta, \delta)$ distribution and added to the current value of the parameter; $\delta$ is a tuning parameter that can be modified to increase the efficiency of the MCMC sampling. The proposal ratio for the sliding window proposal is 1.0, and the acceptance probability is $\min(1, L)$.

275 I placed a uniform $(0, T)$ prior density on the location of transitions, assuming simply that all positions on the tree are equiprobable. Thus, during the addition of a new transition, we sample a new map location at random from $(0, T)$. I placed relatively flat exponential priors on $\lambda$ and $\mu$ and a normal (mean = 0; variance = 0.05) prior on $z$; the latter choice was motivated by the fact that $z = 0$ corresponds to a constant-speciation diversification

280 process (equation 1). I placed an exponential prior on $\Lambda$, the parameter of the Poisson distribution that serves as a prior on the number of transitions on the tree. Larger values for the rate parameter of this exponential distribution imply a greater number of transitions on the tree. I denote this exponential prior on the number of transitions by γ.

285 Likelihoods were computed on branches using a discretization of the constant-rate birth death model that enabled us to approximate time-dependent and diversity-dependent rate variation. Following the notation from Maddison et al. [23], let $D(t)$ represent the probability that some lineage at time $t$ evolves into a clade identical to the observed descendant clade, and let $E(t)$ represent the probability that the lineage goes extinct

290 before the present. Following [43], let $t_N$ be the initial time for such an interval, in units of time before the present, and let $t$ be some earlier time (closer to the root), such that $t > t_N > 0$. It is straightforward to write down the change in $D$ and $E$ as a function of time, such that

$$\frac{dD}{dt} = -(\lambda + \mu)D(t) - 2\lambda D(t)E(t) \qquad (10)$$

295 and

$$\frac{dE}{dt} = \mu - (\lambda + \mu)E(t) + \lambda E(t)^2 \qquad (11)$$

Let $E_0$ and $D_0$ denote the initial values of the speciation and extinction probability for a given interval $\Delta t$ over which $E(t)$ and $D(t)$ must be computed. The analytical solution to equation (11), given $E_0$, is

300 $$E(t) = 1 - \frac{(1 - E_0)(\lambda - \mu)}{(1 - E_0)\lambda - e^{-(\lambda - \mu)t}(\mu - \lambda E_0)} \qquad (12)$$

which is identical to equation (12a) from FitzJohn et al. (2009) under the substitution $E_0$

= 1 - *f*, where *f* is the sampling fraction of the phylogeny. FitzJohn et al. [43]

demonstrated how the birth-death model could be extended to account for incomplete

taxon sampling (via the sampling fraction *f*), but their results allow the calculation of

305    probabilities along any segment of a branch of a phylogenetic tree, provided that $E_0$ and

$D_0$ are available for the beginning of the interval over which the probabilities are to be

computed. Equation (12) can be substituted into equation (10), and the resulting

expression simplifies to

$$D(t) = \frac{e^{(\mu-\lambda)\Delta t} D_0 (\lambda-\mu)^2}{\left[\lambda - \lambda E_0 + e^{(\mu-\lambda)\Delta t} (E_0 \lambda - \mu)\right]^2} \qquad (13)$$

310    where Δt is the duration of the focal interval, between time $t_N$ and time *t*. As we begin

with known conditions $D_0$ and $E_0$ at the beginning of the focal segment, we can set $t_N = 0$

for the purposes of our calculations. This immediately simplifies equation (13) from [43]

and leads to equation (13) above. As demonstrated by FitzJohn et al. [43], these

calculations reduce to the speciation-extinction model for phylogenetic trees developed

315    earlier [12].

To discretize the rate calculations, I broke each branch of the phylogenetic tree into

segments and computed the mean speciation rate under the exponential change model

(equation 1) for the corresponding process. I then assumed constant rate diversification

320    within each branch segment. For each branch segment, the initial speciation and

extinction rates $D_0$ and $E_0$ are equal to the terminal values for the preceding segment. I

made this design choice to facilitate rapid likelihood calculations on large phylogenetic trees and, as demonstrated below, this discretization performs well across a range of simulated datasets. I used a step size of 1.0 time units for all calculations. If a branch was

325    particularly short, such that this step size exceeded the length of the branch, the entire branch was assumed to have a single rate equal to the mean rate along its length.

To compute the likelihood of the full tree under a given set of parameters, we perform the

calculations described above on each terminal branch of the phylogeny. Initial values of

330    $D_0$ and $E_0$ at the tips of the tree were set to 1.0 and 0.0 respectively. It is straightforward

to modify these values to account for incomplete taxon sampling if only a fraction $f$ of the

total species in a clade have been included in a phylogenetic tree. When $f > 0$, we can set

$D_0 = f$ and $E_0 = 1 - f$ for the initial calculations at the tips of the tree [43]. This correction

assumes that species are missing at random from the phylogeny, which may not be valid

335    for many datasets [44,45,46,47]. As in the BiSSE calculations [23], these calculations flow "rootwards" from these terminal branches towards the root. When terminal probabilities have been computed for both descendant branches from a given internal node, the left ($D_L$) and right ($D_R$) branch probabilities were combined as $\lambda\, D_R(t)D_L(t)$,

where $\lambda$ is the speciation rate at the focal node [23]. The calculations then continue down

340    the branch subtending this node. The likelihood of the full tree is the value of $D$ after combining these probabilities at the root node. Likelihoods were conditioned on the occurrence of a root node and on the survival of both descendent branches from the root speciation event [12,23].

345    I implemented the compound Poisson process model of rate variation described above in a C++ program, which I refer to as BAMM.  BAMM (Bayesian Analysis of Macroevolutionary Mixtures) can estimate the number of distinct evolutionary regimes across phylogenetic trees and estimates marginal distributions of speciation and extinction rates for each branch in a phylogenetic tree. The model allows extinction rates

350 to exceed speciation rates. BAMM and associated documentation is available from the BAMM project website (www.bamm-project.org). The program operates on fully bifurcating phylogenetic trees of extant species. The implementation allows users to analytically account for incomplete taxon under the assumption of random taxon sampling [43].

355

Analysis of simulated datasets

To evaluate performance of the compound Poisson process model of diversification rate

360 variation, I simulated phylogenetic trees under six general diversification models. I first considered a simple constant-rate birth death process (model CR; 1 process), to evaluate parameter bias and the frequency of overfitting when the generating model does not include a heterogeneous mixture of processes. Given the widespread interest in identifying well-supported rate shifts and key innovations on phylogenetic trees, we are

365 particularly interested in the frequency with which the model described here will incorrectly identify a multi-process model as having the maximum *a posteriori* probability, when the true generating model is a single process model. To assess whether my results were sensitive to choice of prior on the Poisson rate parameter $\Lambda$, I analyzed constant-rate phylogenies under three different prior parameterizations, corresponding to 370     $\gamma = 1$, $\gamma = 5$, and $\gamma = 10$. All other analyses used a prior of $\gamma = 1.0$, which is conservative in the context of these analyses (see results).

I also considered a model where a pure-birth diversification process shifts to an

exponential change process at some point in time (model exp2; 2 processes). Finally, I

375     considered four variants of diversity-dependent multi-process models. In each case, I

assumed that a pure-birth process at the root of the tree underwent multiple (1, 2, 3, or 4)

shifts to independent and decoupled diversity-dependent speciation-extinction processes

(models DD2, DD3, DD4, DD5). I conducted 500 simulations per scenario.

380     Each multiprocess simulation was conducted by first simulating a pure-birth phylogeny

for 100 time units with $\lambda = 0.032$. I then randomly chose a time $T_s$ on the interval (40, 95)

for the occurrence of a rate shift. A shift was then assigned randomly to one of the

lineages that existed at time $T_s$. I then sampled parameters for the new process (see

below). The tree was then broken at the shift point, and a new subtree was simulated

385     forward in time from the shift point under the new process parameters. For trees with more than two processes, this procedure was repeated until the target number of processes had been added. For the exp2 model, this consisted of sampling $\lambda$, $z$, and $\mu$ for the shift process uniformly on the following intervals: $\lambda$, (0.05, 0.50); $z$, (-0.10, 0.05); $\mu$,

(0.0, 0.45). Thus, the addition of an exponential change process could have resulted in

390     either an increase in rates through time (if $z > 0$) or a decrease (if $z < 0$). For all simulations, I required that subtrees contained at least 25 and fewer than 1000 terminal taxa; any simulations failing to meet this criterion were automatically rejected.

For the diversity-dependent models, diversification dynamics followed a linear diversity-

395    dependent model [48]. The rate of speciation was thus a function of the number of coeval

lineages in the subclade, or

$$\lambda(n) = \lambda_0 \left(1 - \frac{n_t}{K}\right) \qquad (14)$$

400    where $K$ is the clade-specific carrying capacity, and $n_t$ is the number of lineages in the

subclade at time $t$. Note that the occurrence of a shift event results in a decoupling of

dynamics from the parent process. To parameterize the diversity-dependent processes, I

sampled $\lambda_0$ from a uniform (0.05, 0.40) distribution, $K$ from a uniform (25, 250)

distribution, and $\mu$ from a uniform (0, 0.05) distribution. For the constant-rate birth-death

405    simulations, I sampled $\lambda$ from a uniform (0, 0.1) distribution and chose a corresponding relative extinction rate ($\mu / \lambda$) from a uniform (0, 0.99) distribution.

Each of the 500 simulations for each of 6 simulation scenarios was thus conducted under

a potentially unique speciation-extinction parameterization. The number of taxa in each

410    simulated tree also varied among datasets. I recorded the mean rate of speciation and

extinction across each branch in each simulated tree. All simulations were conducted in

C++; simulated trees are available through the Dryad data repository

(doi:10.5061/dryad.hn1vn).

415    I analyzed each of the 3000 simulated datasets using BAMM with 3 million generations of MCMC sampling. I discarded the first half of samples from each simulation of the posterior as "burn-in" and estimated the overall "best model" as the model that was sampled most frequently by the Markov chain. I computed the mean of the posterior distribution of speciation and extinction rates on each branch for each tree. I then used

420    OLS regression to assess the relationship between branch-specific rate estimates obtained using BAMM versus the true underlying evolutionary rates. As an additional estimate of bias, I computed the proportional error [37] in the estimated rates as a function of the true rates. This metric is computed as the weighted average of proportional rate differences across all $N$ branches in the phylogeny, or

425

$$PE = \exp\left(\frac{1}{N}\sum\Big[\log\big(r_{EST}\big) - \log\big(r_{TRUE}\big)\Big]\right) \qquad (15)$$

where $r_{EST}$ and $r_{TRUE}$ are the estimated and true values of rates along a particular branch. A value of 2 would imply that estimated rates are, on average, equal to twice the true rate

430    in the generating model.

Comparison with MEDUSA

I compared the performance of BAMM to that of MEDUSA [20], a maximum likelihood

435    method for modeling among-lineage heterogeneity in speciation-extinction dynamics. Beginning with a constant rate birth-death process, MEDUSA uses a stepwise AIC algorithm to incrementally add rate shifts to phylogenetic trees until the addition of new partitions fails to improve the fit of the model to the data. Thus, MEDUSA is similar to the method described here in that it is explicitly designed to discover the number and
location of distinct processes of speciation and extinction on phylogenetic trees. However,
MEDUSA, as implemented and typically used, makes the assumption that rates of species
diversification are constant in time within rate classes. This assumption has been rejected
by studies across a range of taxonomic scales, from species-level phylogenies
[16,17,18,26] to tree-of-life scale compilations of clade age and species richness [49].
However, the consequences of violating this assumption for MEDUSA analyses have not
been investigated.

I analyzed each of the simulated datasets described above (500 datasets under each of 6
distinct models of diversification) using MEDUSA, using the implementation of
MEDUSA available in the Geiger v1.99-3 package [50] for the R programming and
statistical environment. Model selection used the default AICc criterion. I summarized
the results of MEDUSA analyses in two ways. First, for each simulation scenario, I
tabulated the distribution of "best fit" models, to assess the fraction of simulations for
which MEDUSA was able to correctly estimate the number of processes in the generating
model. Second, I used the same summary statistics described above for BAMM (e.g.,
proportional error) to compare branch-specific estimates of speciation rates under
MEDUSA to the true rates under the generating model.

Empirical example: cetacean radiation

Steeman et al. [51] provided a time-calibrated phylogenetic tree for 87 of 89 extant species of whales and dolphins (Mammalia: Cetacea). They found support for increased

rates of species diversification within a major dolphin clade, the Delphinidae. They also

found evidence for an increased rate of species diversification at approximately 7.5

465    million years before present (Ma). I used the BAMM implementation of the compound

Poisson process model of diversification rate variation to investigate the tempo and mode

of cetacean diversification through time. I conducted 5 million generations of MCMC

sampling, with multiple independent runs to assess convergence. Finally, I assessed the

sensitivity of the cetacean analyses to the choice of prior $\gamma$ on the number of processes in

470    the phylogeny.

Results

Analysis of simulated datasets

475

Figure 2 shows a representative BAMM analysis for a tree simulated under the DD3

model (see also Figure 1). BAMM results are generally robust to choice of prior on the

expected number of processes ($\gamma$) in the phylogeny under the compound Poisson process

model of rate variation (Figure 3). With increasing values of $\gamma$, the model with maximum

480    *a posteriori* probability (MAP) was biased in favor of $M_1$, a model with two processes.

However, this represents a weak tendency towards model overfitting, because the true

model ($M_0$) was generally characterized by a posterior probability much greater than 0.05

(Figure 3E, F). This suggests that results are robust to choice of $\gamma$: even with the trend

towards overfitting (Figure 3C), the method is unlikely to yield strong support for models

485    that are more complex than the generating model (Figure 3F). The simulation results presented below are based on $\gamma = 1$.

For the five simulation scenarios with rate heterogeneity, the number of distinct processes

estimated using BAMM was generally equal to the number of processes in the generating

490    model (Figure 4). Power to infer the true number of processes decreased for the most

complex models (DD3, DD4, DD5), but model overfitting was not a problem. The MAP

model was no more complex than the generating model in the overwhelming majority of simulations (> 95%) for all five simulation scenarios.

495    Estimates of speciation and extinction rates under the constant-rate model were highly correlated with rates in the generating model (Figure 5), although both rates were biased upwards for low rates. For multiprocess simulation models, branch-specific estimates of speciation rates were highly correlated with rates in the generating model (Figure 6, left).

The estimated slope of the relationship between the true rates and estimated rates

500    approached equality. However, a small percentage of simulations had estimated slopes

that suggested a lack of relationship between true and estimated rates. These simulations

were those where the most frequently sampled model had only a single process and thus

reflect a lack of power, rather than consistent bias. In other words, branch specific

estimates of rates for a multiprocess model may be poor if model underfitting has

505    occurred. In the extreme case, a tree that is estimated to have only a single process may have very similar rate estimates on each branch; the correlation between these rates and the true rates will necessarily be low if the true model includes multiple processes and considerable rate heterogeneity across the tree.

510    A large fraction of the total variation in the underlying speciation rates is also explained by the estimated rates under the compound Poisson process model of diversification rate variation (Figure 6, middle). This fraction is higher for the two-process models (exp2,

DD2) but remains stable across the remaining diversity-dependent scenarios (DD3, DD4,

DD5). Finally, analysis of proportional error suggests that, on average, rates are not

515    consistently over- or underestimated using BAMM; the mean proportional error between simulated and estimated rates is near 1.0 for all simulation scenarios.

I performed a similar analysis for extinction rates, with one key difference. Each

simulation scenario assumed constant extinction rates within each process; hence, the

520    number of unique extinction rates in each simulation was equal to the number of processes in the generating model. I computed the mean branch-specific extinction rate across each subclade that was governed by a distinct evolutionary process in the simulation model; I then analyzed these extinction rates across all 500 trees in a given simulation scenario together. For example, consider the phylogeny shown in Figure 1,

525    with $k = 3$ processes. I computed the mean of the posterior distribution of all extinction

estimates for branches assigned to process $A$, giving a single estimated rate overall for

that process. I repeated this for processes $B$ and $C$, such that I obtained $k$ extinction

estimates for each tree with $k$ processes. Table 1 shows summary statistics for analyzing

these sets of extinction rate estimates across all trees from a given simulation scenario. In

530    general, relative rate differences suggest that extinction estimates are biased upwards.

Nonetheless, the fraction of variance explained by the model is low in each case. Proportional error calculations were also performed as described above, although the root rate class was ignored, as it was equal to 0 in the simulation model. These values likewise indicate an upwards bias in extinction rate estimates.

535

Comparison with MEDUSA

I analyzed all 3000 simulated datasets using MEDUSA. The number of processes inferred for each simulation scenario are shown in Figure 7 and can be compared to the corresponding BAMM results shown in Figure 3. MEDUSA performed worse than BAMM for all scenarios with time-varying rates of species diversification. Under the DD2 model, with just two processes, MEDUSA estimated the correct number of processes in 40.2% of simulated datasets. In contrast, BAMM correctly identified the true number of processes in 90% of simulated datasets (Figure 3). For diversity-dependent scenarios with more than two processes in total, MEDUSA consistently underestimated the true number of processes in the generating model for the overwhelming majority of simulated datasets. For the DD5 scenario, MEDUSA correctly identified the generating model in fewer than 5% of simulations, versus 38.4% with BAMM (Figure 7 vs. Figure 4).

Branch-specific estimates of speciation rates under MEDUSA were, in general, extremely poor (Figure 8) when compared to the corresponding estimates under BAMM (Figure 6). The estimated slope of the relationship between the true speciation rates on each branch and the corresponding MEDUSA estimates has a modal value of zero for four of five simulation scenarios (Figure 8, left column). In contrast, BAMM estimates were far

555    closer to the 'perfect' value of 1.0. For all simulation scenarios but exp2, the MEDUSA-

estimated speciation rates explained little of the variance in the true distribution of rates

(Figure 8, middle). Finally, proportional error analysis indicated that MEDUSA generally underestimates true rates of speciation when rates are time-dependent or diversity-dependent (Figure 8, right column).

560

Empirical example: cetacean radiation

Analysis of the time-calibrated cetacean phylogeny found strong support for a two-process model (Figure 9). The posterior probability of a one-process model is $p = 0.017$,

565    with a posterior odds ratio of 44.6 in favor of a two-process model. These results suggest

a substantial increase in the rate of speciation in the ancestral lineage leading to the

Delphinidae (Figure 9A), possibly excluding the killer whale *Orcinus orca*. The posterior

probability of a rate shift occurring on at least one of these branches is greater than 0.975

(Figure 9B). However, we find little evidence for additional processes within the cetacean

570    phylogeny as a whole (Figure 9 B, C).

Using output from BAMM, I computed mean rates of speciation and extinction through

time during the cetacean radiation. This was done by drawing an imaginary grid of

vertical lines through the time-calibrated cetacean phylogeny at equally spaced points in

575    time. Evolutionary rates were estimated as the mean branch-specific rates for all branches that intersected the line corresponding to a specific time point. This enabled estimation of the posterior density of speciation and extinction for any point in time. These results

suggest an overall decline in the background rate of whale speciation, with a large spike

during the Miocene driven by the radiation of the dolphin clade (Delphinidae). Extinction

580    rates are inferred to be relatively low overall, with a mean per-branch relative extinction rate ($\mu$ / $\lambda$) of 0.36.

Finally, I assessed the sensitivity of the cetacean dataset to choice of prior on the number

of processes in the phylogeny. I used BAMM to analyze the cetacean data under four

585    additional values of $\gamma$ (0.1, 0.5, 5, and 10). In each case, the single-process model had low posterior probability and was marginally worth considering (Pr ($M_1$) = 0.127) only under the strongest prior ($\gamma$ = 0.1). For $\gamma$ = 5 and $\gamma$ = 10, the posterior probability of a model

with a single process was approximately 0. The MAP model had two processes under $\gamma$ =

0.5 and $\gamma$ = 5. For $\gamma$ = 10, the MAP model had four processes, but was not substantially

590    more probable than models with two or three processes. The posterior odds ratio for $M_4$

versus $M_2$ was merely 1.63, and for $M_4$ versus $M_3$ it was only 1.26. I estimated marginal

diversification rates for each branch in the cetacean phylogeny under these prior

formulations; pairwise plots for speciation rate estimates under alternative priors suggest

that these rates are robust to choice of prior (Figure 10). Extinction rate estimates were

595    sensitive to choice of prior, although estimated rates were low under all prior formulations.

Discussion

600     Extracting information about the tempo and mode of species diversification remains a central methodological challenge in macroevolutionary studies. I developed a Poisson process model of diversification rate variation to address several limitations of current methodological approaches for studying evolutionary dynamics on phylogenetic trees.

605     The approach described here views phylogenetic trees as the outcome of a complex

mixture of potentially dynamic evolutionary processes and enables researchers to detect

rate shifts, key innovations, time-dependent speciation, and diversity-dependence within

single trees. Output from the BAMM implementation of the compound Poisson process

model includes (i) estimates of the number of distinct process and posterior probabilities

610     of each possible model; (ii) estimates of locations of those processes as well as associated

parameter estimates; and (iii) estimates of branch-specific rates of speciation and

extinction, which can further be used to infer temporal trends in evolutionary rates

(Figure 9D, E).

615     BAMM performed well throughout the parameter space explored here. For each of six

distinct macroevolutionary scenarios, BAMM was usually able to identify the true

number of processes in the generating model (Figure 3; Fig.4). Branch-specific speciation

rates estimated using BAMM are fairly accurate: relative rate differences for estimated

rates are centered on 1 (Figure 6, right). Moreover, the OLS regression slope for the

620     relationship between true and estimated branch-specific rates across individual simulation

trees was generally close to 1.0; the mean of each distribution of slopes shown in Figure

6 (left column) exceeded 0.85. Surprisingly, branch-specific estimates did not decay with increasing complexity of the generating model: observed slopes (Figure 6, left) for the most complex model (DD5) were closer to 1.0 (observed mean: 0.95) and had lower variance than any other simulation scenario, including those with only two processes.

Extinction rate estimates from the model should be taken with caution. Branch specific estimates of extinction are potentially biased and, although these estimates are correlated with the true underlying rates, confidence in those estimates is low (Table 1; Figure 5). This is consistent with previous studies that have noted low power in estimating extinction rates from molecular phylogenies [23,52]. In addition, previous studies have demonstrated that extinction estimates from molecular phylogenies are exceedingly sensitive to violations of model assumptions [27,53]. Because few real-world phylogenies will conform perfectly to the assumptions of the model described here, it is likely that estimated extinction rates will be even less accurate than results in Table 1 would suggest.

By implementing an exponential change function for speciation, I was able to accurately infer diversity-dependent dynamics across a range of simulation scenarios (Figure 4; Figure 6). This is consistent with Quental and Marshall's [34] prediction that time-dependent exponential processes (equation 1) should provide good approximations to linear diversity-dependent processes. It is possible that formal diversity-dependent models [13,48,54] would provide increased power and/or precision of parameter estimates over the exponential approximation used in this study. However, fitting a full diversity-dependent model with extinction is far more computationally demanding than the exponential approximation used here. For multiprocess diversity-dependent models, computing a single likelihood currently requires numerically solving large but linear systems of ordinary differential equations. The exponential approximation implemented in BAMM results in extremely fast likelihood calculations on even the largest phylogenetic trees. No attempts have yet been made to parallelize BAMM calculations, affording additional opportunities for computational speedups.

Comparison to existing methods

My results suggest that MEDUSA is not robust to violations of its assumption that diversification rates are constant through time. Whereas BAMM was often able to estimate the true number of distinct processes in the generating model (Figure 4), MEDUSA consistently underestimated the number of processes (Figure 7). Furthermore, the magnitude of the underestimates became more severe with increasing model complexity. Speciation rates estimated under MEDUSA were especially poor (Figure 8) and showed little overall correspondence with true rates in the simulation model.

To be clear, the model implementation in BAMM - in contrast to MEDUSA - was explicitly designed to account for variation in evolutionary rates both through time and among lineages. However, the MEDUSA method has been applied to many empirical datasets with little attention given to the potential consequences of violating the assumption of rate-constancy through time. Using *a posteriori* simulations, Rabosky et al. [49] found that parameter estimates from MEDUSA analyses on higher taxonomic datasets largely fail to predict patterns of species richness across clades. They suggested that this failure results from MEDUSA's strong assumption of time-invariant speciation

670    and extinction rates. It seems likely that many or most real datasets will be characterized by rate variation through time as well as among lineages. As discussed by O'Meara [55],

the challenges of modeling rate heterogeneity in phylogenetic trees are likely to become

more severe as we consider ever-larger phylogenetic trees [56,57,58,59]: the larger the

phylogeny, the greater the likelihood that the tree is the result of a heterogeneous mixture

675    of distinct evolutionary processes. Describing the complex mixture of dynamic processes that shape real phylogenetic trees was the primary motivation for proposing the method described in this article.

Cetacean macroevolutionary dynamics

680

The analysis of the Cetacean phylogeny provides an important window into the history of cetacean diversification through time (Figure 9) that complements results obtained by several previous studies [6,13,51]. The overall lineage accumulation curve for cetaceans is relatively flat [13], suggesting relatively little variation in speciation rates through time.

685    However, I find strong support for a multi-process diversification model consisting of

two distinct evolutionary rate regimes: a root process involving a weak slowdown in

speciation through time (Figure 9A), and an explosive burst and subsequent slowdown in

speciation associated with the origin of the Delphinidae (Figure 9A). Slater et al. [60]

also found support for a rate shift in the crown delphinids, excluding the killer whale,

690    using MEDUSA. It seems likely that some of the evidence in favor of the "ocean restructuring" model [51] actually reflects the independent evolutionary dynamics of delphinid and non-delphinid lineages. The increase in speciation from 13 million years ago (Ma) to 4 Ma in particular seems likely to indicate the rapid diversification of the dolphin clade. My results do not rule out the possibility that ocean restructuring

695    contributed to this clade-specific burst and slowdown in speciation rates, but it appears equally plausible that the acceleration in rates during this interval reflects the occurrence of a key evolutionary innovation early in the history of the dolphins.

Extensions to the model

700

Many extensions are possible within the framework developed here. The computational machinery for adding, moving, and deleting processes from phylogenetic trees is flexible and can easily be extended to allow alternative functional models for speciation and/or extinction rate variation through time. Another obvious future extension is to explicitly

705    account for phylogenetic uncertainty during simulation of the posterior. As currently implemented, BAMM simulates posterior distributions of models and parameters across a fixed topology. However, phylogenetic trees are rarely (if ever) known without error. Credible intervals on parameters inferred using BAMM (Figure 9 D, E) reflect only parametric uncertainty associated with the diversification model itself and would

710    presumably increase if we also accounted for uncertainty in tree topology and branch lengths. Finally, it would be interesting to allow joint inference on paleontological and neontological data, as there is increasing recognition that these two datatypes are frequently in conflict [61]. This objective is facilitated by theoretical advances that allow evolutionary rate estimation using both fossils and molecular phylogenies [62], although 715    suitable datasets remain elusive.

Summary

I have described a methodological framework for inferring mixtures of processes that

720    have influenced the structure of phylogenetic trees. By modeling phylogenies as

collections of dynamic processes, the method greatly extends our ability to describe

evolutionary dynamics. Most previous evolutionary studies using transdimensional

MCMC on phylogenetic trees have assumed that dynamics within component processes are constant in time. By relaxing the assumption of time-homogeneous diversification,

725    the model is better able to describe complex mixtures of both time-constant and time-

varying processes. A number of recent studies have suggested that such complex

dynamics might dominate speciation-extinction patterns in many empirical datasets. I

suggest that the use of rjMCMC to fit time-inhomogeneous multiprocess models to

phylogenetic data may have applications beyond those described here, including DNA

730    sequence evolution, phenotypic evolution, and phylogeography.

Acknowledgements

I thank M. Alfaro. F. Bokma, R. FitzJohn, L. Harmon, J. Huelsenbeck, S. Kolokotronis,

735    B. Moore, S. Otto, A. Pyron, J. Schraiber, J. Uyeda, and members of the Huelsenbeck lab

for comments on the manuscript and/or discussion that improved the results presented here. I thank T. Nan for developing the PPSS code that facilitated parallel analysis of simulated datasets on multiprocessor computers. Financial support was provided by the

Miller Institute for Basic Research in Science at the University of California, Berkeley,

740    by the University of Michigan, and by NSF- DEB-1256330.

References

1. Alroy J (2010) The shifting balance of diversity among major marine animal groups. Science 321: 1191-1194.

745   2. Sepkoski JJ (1978) A kinetic model of Phanerozoic taxonomic diversity I. Analysis of marine orders. Paleobiology 4: 223-251.

3. Mittelbach GG, Schemske DW, Cornell HV, Allen AP, Brown JM, et al. (2007) Evolution and the latitudinal diversity gradient: speciation, extinction and biogeography. Ecology Letters 10: 315-331.

750   4. Ricklefs RE (1987) Community diversity - relative roles of local and regional processes. Science 235: 167-171.

5. Glor RE (2010) Phylogenetic insights on adaptive radiation. Ann Rev Ecol Evol Syst 41: 251-270.

6. Morlon H, Parsons TL, Plotkin JB (2011) Reconciling molecular phylogenies with the

755   fossil record. Proceedings of the National Academy of Sciences of the United States of America 108: 16327-16332.

7. Paradis E (2011) Time-dependent speciation and extinction from phylogenies: a least squares approach. Evolution 65: 661-672.

8. Ricklefs RE (2007) Estimating diversification rates from phylogenetic information.

760   Trends in Ecology & Evolution 22: 601-610.

9. Wagner CE, Harmon LJ, Seehausen O (2012) Ecological opportunity and sexual selection together predict adaptive radiation. Nature 487: 366-369.

10. Alroy J (2008) The dynamics of origination and extinction in the marine fossil record. Proc Nat Acad Sci USA 105: 11536-11542.

765    11. Foote M (2000) Origination and extinction components of taxonomic diversity:

general problems. Paleobiology 26: 74-102.

12. Nee S, May RM, Harvey PH (1994) The reconstructed evolutionary process. Phil

Trans R Soc Lond B Biol Sci 344: 305-311.

13. Etienne RS, Haegeman B, Stadler T, Aze T, Pearson PN, et al. (2012) Diversity-

770    dependence brings molecular phylogenies closer to agreement with the fossil record. Proc R Soc Lond B Biol Sci 279: 1300-1309.

14. Rabosky DL, Lovette IJ (2008) Explosive evolutionary radiations: Decreasing

speciation or increasing extinction through time? Evolution 62: 1866-1875.

15. Stadler T (2011) Mammalian phylogeny reveals recent diversification rate shifts. Proc

775    Natl Acad Sci USA 108: 6187-6192.

16. McPeek MA (2008) Ecological dynamics of clade diversification and community

assembly. Am Nat 172: E270-E284.

17. Morlon H, Potts MD, Plotkin JB (2010) Inferring the Dynamics of Diversification: A

Coalescent Approach. PLoS Biology 8: e1000493.

780    18. Phillimore AB, Price TD (2008) Density dependent cladogenesis in birds. Plos

Biology 6: e71.

19. Ruber L, Zardoya R (2005) Rapid cladogenesis in marine fishes revisited. Evolution

59: 1119-1127.

20. Alfaro ME, Santini F, Brock C, Alamillo H, Dornburg A, et al. (2009) Nine

785    exceptional radiations plus high turnover explain species diversity in jawed vertebrates. Proc Nat Acad Sci USA 106: 13410-13414.

21. Slowinski JB, Guyer CG (1989) Testing the stochasticity of patterns of organismal diversity: an imporoved null model. American Naturalist 134: 907-921.

22. FitzJohn R (2010) Quantitative traits and diversification. Syst Biol 59: 619-633.

790     23. Maddison WP, Midford PE, Otto SP (2007) Estimating a binary character's effect on speciation and extinction. Syst Biol 56: 701-710.

24. Mitter C, Farrell B, Wiegmann B (1988) The phylogenetic study of adaptive zones - has phytophagy promoted insect diversification? American Naturalist 132: 107-128.

795     25. Paradis E (2005) Statistical analysis of diversification with species traits. Evolution 59: 1-12.

26. Rabosky DL, Glor RE (2010) Equilibrium speciation dynamics in a model adaptive radiation of island lizards. Proc Nat Acad Sci USA 107: 22178-22183.

27. Rabosky DL (2010) Extinction rates should not be estimated from molecular

800     phylogenies. Evolution 64: 1816-1824.

28. Rabosky DL (2012) Testing the time-for-speciation effect in the assembly of regional biotas. Methods Ecol Evol 3: 224-233.

29. Jonnson KA, Fabre P-H, Fritz SA, Etienne RS, Ricklefs RE, et al. (2012) Ecological and evolutionary determinants for the adaptive radiation of the Madagascan

805     vangas. Proc Natl Acad Sci USA 109: 6620-6625.

30. Huelsenbeck JP, Larget B, Swofford D (2000) A compound Poisson process for relaxing the molecular clock. Genetics 154: 1879-1892.

31. Green PJ (1995) Reversible jump Markov chain Monte Carlo computation and Bayesian model determination. Biometrika 82: 711-732.

810    32. Rabosky DL, Donnellan SC, Talaba AL, Lovette IJ (2007) Exceptional among-lineage variation in diversification rates during the radiation of Australia's most diverse vertebrate clade. Proc R Soc Lond B Biol Sci 274: 2915-2923.

33. Bartolucci F, Scaccia L, Mira A (2006) Efficient Bayes factor estimation from the reversible jump output. Biometrika 93: 41-52.

815    34. Quental TB, Marshall CR (2009) Extinction during evolutionary radiations: reconciling the fossil record with molecular phylogenies. Evolution 63: 3158-3167.

35. Bokma F (2008) Detection of "punctuated equilibrium" by Bayesian estimation of speciation and extinction rates, ancestral character states, and rates of anagenetic

820    and cladogenetic evolution on a molecular phylogeny. Evolution 62: 2718-2726.

36. Silvestro D, Schnitzler J, Zizka G (2011) A Bayesian framework to estimate diversification rates and their variation through time and space. BMC Evol Biol 11: 311.

37. Eastman JM, Alfaro ME, Joyce P, Hipp AL, Harmon LJ (2011) A novel comparative

825    method for identifying shifts in the rate of character evolution on trees. Evolution 65: 3578-3589.

38. Venditti C, Meade A, Pagel M (2011) Multiple routes to mammalian diversity. Nature 479: 393-396.

39. Hastings WK (1970) Monte Carlo sampling methods using Markov chains and their

830    applications. Biometrika 57: 97-109.

40. Metropolis N, Rosenbluth AW, Rosenbluth MN, Teller AH, Teller E (1953) Equations of state calculations by fast computing machines. J Chem Phys 21: 1087-1091.

41. Sisson SA (2005) Transdimensional Markov Chains. J Amer Statist Assoc 100: 1077-1089.

42. Yang Z (2006) Computational Molecular Evolution: Oxford University Press.

43. FitzJohn R, Maddison WP, Otto SP (2009) Estimating trait-dependent speciation and extinction rates from incompletely resolved phylogenies. Syst Biol 58: 595-611.

44. Brock CD, Harmon LJ, Alfaro ME (2011) Testing for temporal variation in diversification rates when sampling is incomplete and nonrandom. Syst Biol 60: 410-419.

45. Cusimano N, Renner SS (2010) Slowdowns in diversification rates from real phylogenies may not be real. Systematic Biology 59: 458-464.

46. Hohna S, Stadler T, Ronquist F, Britton T (2011) Inferring speciation and extinction rates under different sampling schemes. Mol Biol Evol 28: 2577-2589.

47. Stadler T (2013) Recovering speciation and extinction dynamics based on phylogenies. J Evol Biol doi: 10.1111/jeb.12139.

48. Rabosky DL, Lovette IJ (2008) Density-dependent diversification in North American wood warblers. Proc R Soc Lond B Biol Sci 275: 2363-2371.

49. Rabosky DL, Slater GJ, Alfaro ME (2012) Clade age and species richness are decoupled across the Eukaryotic tree of life. PLoS Biol 10: e1001381.

50. Harmon LJ, Weir JT, Brock C, Glor RE, Challenger WE (2008) GEIGER: Investigating evolutionary radiations. Bioinformatics 24: 129-131.

51. Steeman ME, Hebsgaard MB, Fordyce RE, Ho SYW, Rabosky DL, et al. (2009) Radiation of Extant Cetaceans Driven by Restructuring of the Oceans. Systematic Biology 58: 573-585.

52. Nee S, Holmes EC, May RM, Harvey PH (1994) Extinction rates can be estimated from molecular phylogenies. Philos Trans R Soc Lond B 344: 77-82.

53. Rabosky DL (2010) Primary controls on species richness in higher taxa. Systematic Biology 59: 634-645.

54. Etienne RS, Haegeman B (2012) A conceptual and statistical Framework for adaptive radiations with a key role for diversity dependence. Am Nat 180: E75-E89.

55. O'Meara BC (2012) Evolutionary inferences from phylogenies: a review of methods. Ann Rev Ecol Evol Syst 43: 267-285.

56. Bininda-Emonds ORP, Cardillo M, Jones KE, MacPhee RDE, Beck RMD, et al. (2007) The delayed rise of present-day mammals. Nature 446: 507-512.

57. Jetz W, Thomas GH, Joy JB, Hartmann K, Mooers A (2012) The global diversity of birds in space and time. Nature 491: 444-448.

58. Pyron RA, Burbrink FT, Wiens JJ (2013) A phylogeny and updated classification of Squamata, including 4161 species of lizards and snakes. BMC Evol Biol 13: 93.

59. Rabosky DL, Santini F, Eastman JM, Smith SA, Sidlauskas B, et al. (2013) Rates of speciation and morphological evolution are correlated across the largest vertebrate radiation. Nature Communications 4: 10.1038/ncomms2958.

60. Slater GJ, Price SA, Santini F, Alfaro ME (2010) Diversity versus disparity and the radiation of modern cetaceans. Proc R Soc Lond B Biol Sci 277: 3097-3104.

61. Quental TB, Marshall CR (2010) Diversity dynamics: molecular phylogenies need the fossil record. Trends Ecol Evol 25: 434-441.

62. Didier G, Royer-Carenzi M, Laurin M (2012) The reconstructed evolutionary process with the fossil record. J Theor Biol 315: 26-37.

880   63. Pybus, OG, Harvey, PH (2010) Testing macro-evolutionary models using incomplete molecular phylogenies. Proc R Soc B 267: 2267 - 2272.

Figures

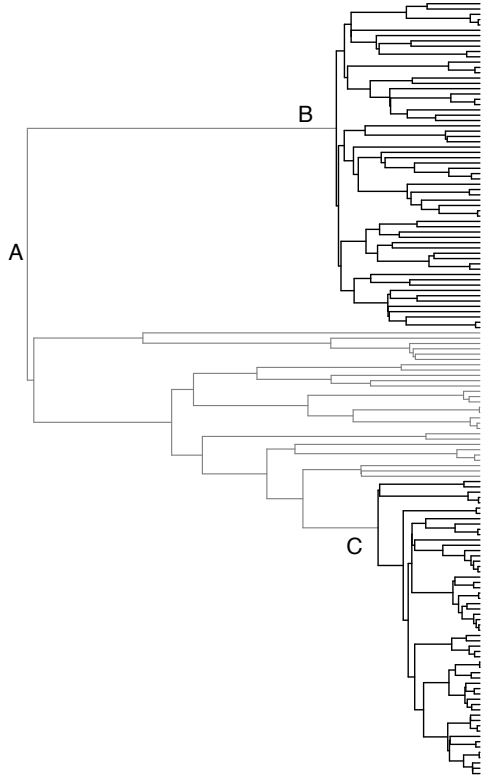

885

**Figure 1. Example of tree simulated under mixture of three distinct evolutionary processes**. (A) Clade diversification under constant-rate "background" diversification process with $\lambda = 0.032$ and $\mu = 0$. (B) Shift to new adaptive zone with subsequent diversity-dependent regulation of speciation and diversity-independent extinction (blue branches; $\lambda_0 = 0.395$; $K = 66$; $\mu = 0.041$). (C) Another lineage shifts to diversity-dependent speciation regime (red branches; $\lambda_0 = 0.21$; $K = 97$; $\mu = 0.012$). Total tree depth is 100 time units. Despite undergoing two distinct diversity-dependent slowdowns in the rate of speciation, the overall gamma statistic [63] for the tree is positive ($\gamma = 2.51$) and provides no evidence for changes in the rate of speciation through time. Note that a tree with three distinct processes contains two distinct transitions between processes.

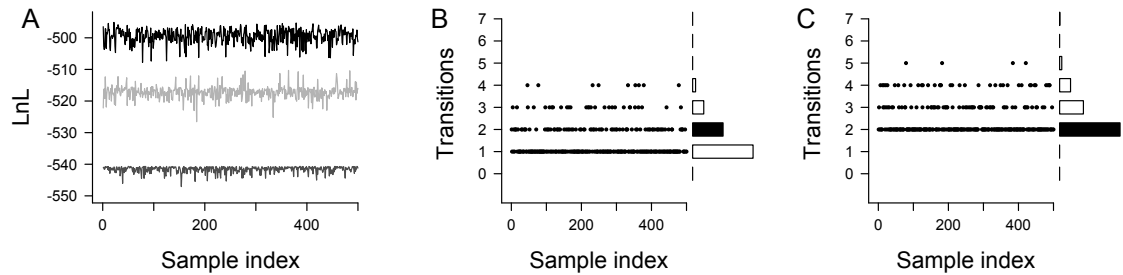

 **Figure 2. BAMM analysis of example tree (Figure 1)**. Example tree was simulated under three distinct processes (one constant rate and two diversity dependent processes; two transitions in total). The tree was analyzed under (i) the full multi-process BAMM model with time-variable speciation; (ii) a constrained multi-process BAMM with time-constant speciation; and (iii) a fully constrained 1-process constant-rate birth-death model.  (A) Log-likelihoods for thinned MCMC chains for the constant rate birth-death process (bottom), the time-constant multi-process model (middle), and the full BAMM model with time-varying speciation (top). (B) Numbers of transitions during rjMCMC sampling when model is constrained to time-constant speciation rates; sidebar gives frequency distribution of sampled states. (C) Numbers of transitions under full BAMM model with  time-variable speciation processes. Black sidebar denotes true number of transitions in generating model. The true number of transitions was estimated correctly only when the assumption of time-constant rates was relaxed.

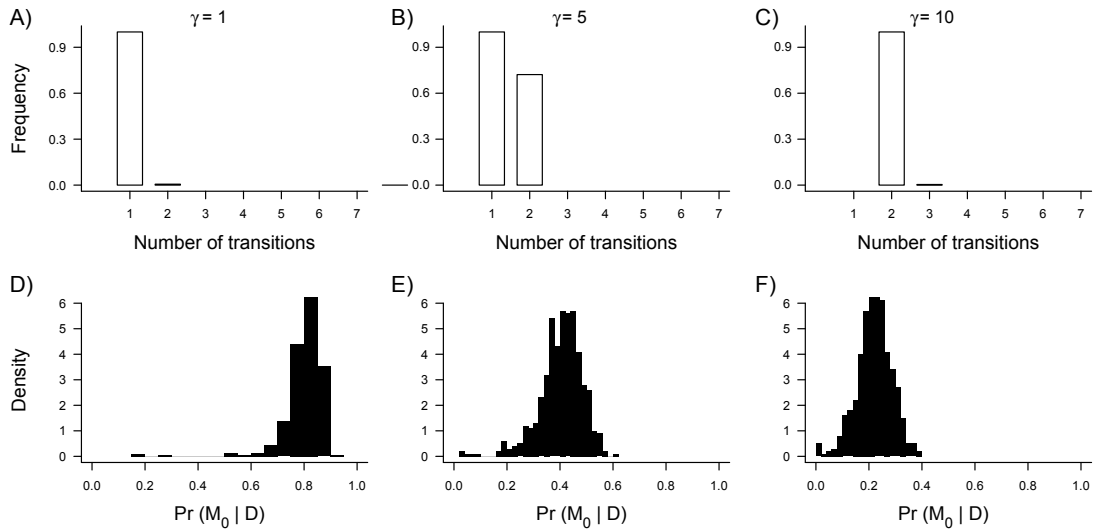

**Figure 3. BAMM analyses of constant-rate phylogenies and prior on Poisson rate parameter (γ).** Histograms in (A-C) display the frequency distribution of the estimated number of processes in the model with the maximum a posteriori (MAP) probability as a function of three different priors on the Poisson rate parameter $\Lambda$ (γ = 1; γ = 5; γ = 10). This "best-fit" model was simply the model that was visited most often during the MCMC simulation of the posterior. (D-F) show the distribution of posterior probabilities for the true model ($M_0$). With a relatively flat prior on models (γ = 10), the MAP model is biased towards a model with 2 processes (= 1 transition). However, the posterior probability of the true model $M_0$ remains substantial (F), and $M_0$ nonetheless had a posterior probability greater than 0.10 for the vast majority of simulations. Results are based on 500 simulated phylogenies per γ scenario.

930

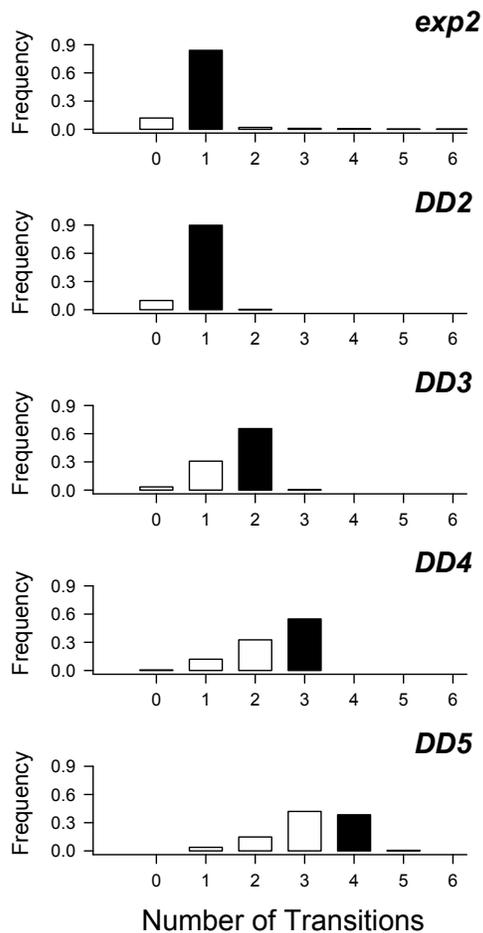

Figure 4. **Frequency distribution of evolutionary rate regimes estimated using BAMM, compared with true number of processes**. For each simulation, the estimated number of processes was simply the model that was most frequently sampled during

935 MCMC simulation of the posterior distribution. Black bars denote true number of processes in generating model. For example, 84% of trees simulated under a single-shift exponential change model (exp2; top panel; two processes in the generating model) were correctly inferred to have been generated under a two-process model. For a diversity-dependent model with five processes (DD5), power to detect the true number of

940 processes is lower, although though most analyses (80.4%) recovered either 4 or 5 process models as the MAP model. Results for each model are based on 500 simulated phylogenies and used a conservative $\gamma = 1$ prior on the expected number of non-root processes (see Figure 3).

945

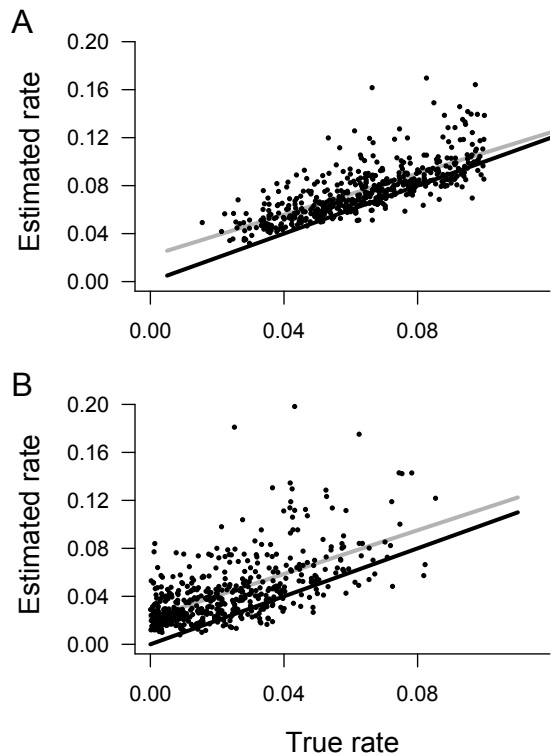

**Figure 5. BAMM estimates of speciation and extinction rates for phylogenies simulated under constant-rate birth-death process.** (A) Relationship between speciation rate in generating model and reconstructed mean rate across the tree under BAMM. Solid black line: identity line, expected if $\lambda_{\text{TRUE}} = \lambda_{\text{ESTIMATED}}$. Solid gray line: fitted OLS regression to estimates (black points) obtained using full BAMM model (multiple processes with time-variable speciation rates). (B) Corresponding extinction rate estimates for same set of trees.

950

955

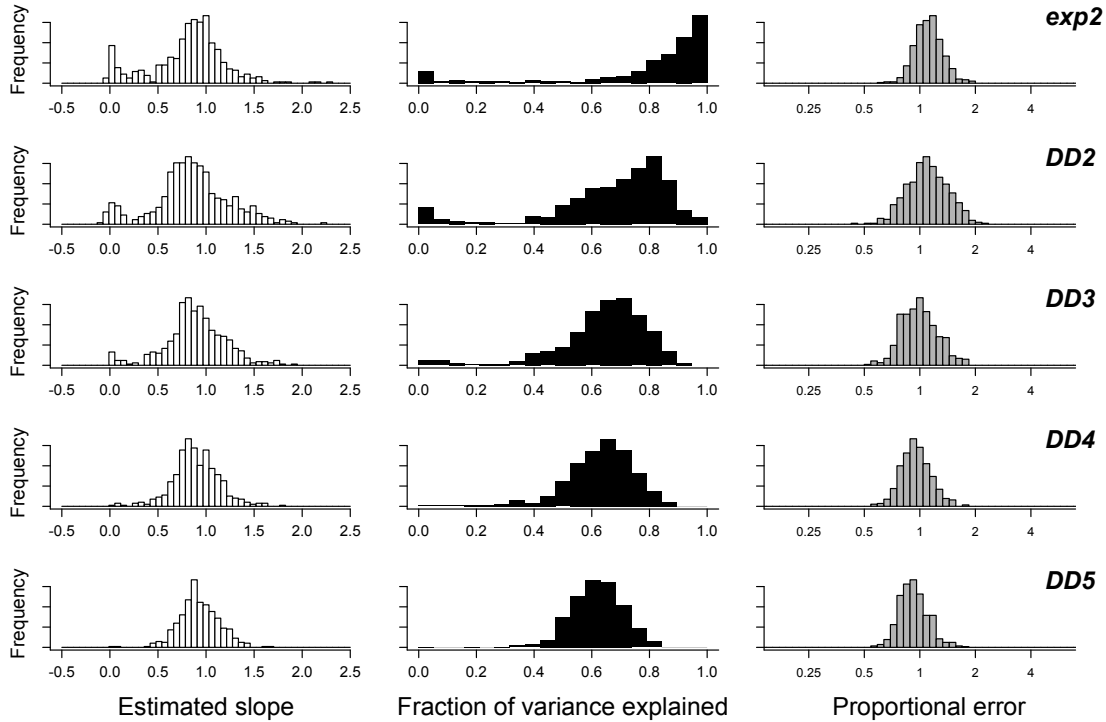

**Figure 6. Precision and bias of BAMM in the estimation of branch-specific rates of speciation.** Phylogenies were simulated under 5 distinct evolutionary scenarios. For each simulated phylogeny, I reconstructed branch-specific speciation rates using BAMM and modeled these as a function of the true branch rates from the generating model. Frequency distributions of the estimated slope of this relationship are shown in the left column for each simulation scenario. Center column denotes corresponding $r^2$ values from the same OLS regressions. Right column is distribution of mean relative rate differences (RRD) for each scenario. A value of 1 implies that, on average, branch-specific speciation estimates are unbiased; a value of 0.5 would imply that branch-specific estimates are, on average, equal to 50% of the true value. Results for each simulation scenario are based on 500 simulated phylogenies (thus giving 500 slopes, $r^2$ values, and RRD values for each simulation scenario).

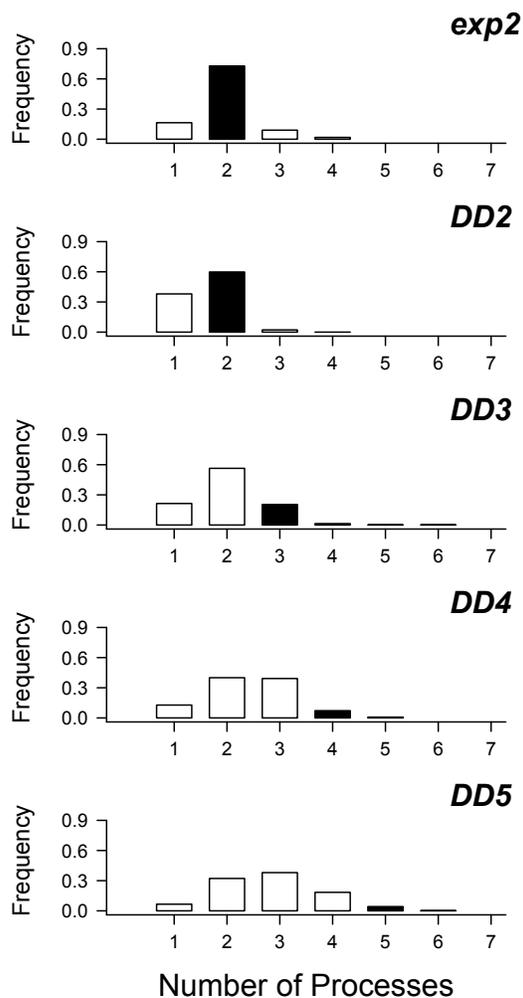

 **Figure 7. Frequency distribution of evolutionary rate regimes estimated using MEDUSA, compared with true number of processes**. Phylogenies were simulated under 5 distinct evolutionary scenarios.  For each simulation, the number of distinct rate partitions was estimated using the stepwise AICc algorithm as implemented in MEDUSA. Black bars denote true number of processes in generating model. MEDUSA consistently  underestimates the true number of processes in simulated datasets when rates of speciation vary through time. Comparable results for BAMM using the same set of simulated datasets are shown in Figure 4. A total of 500 simulated datasets were analyzed per diversification scenario.

985

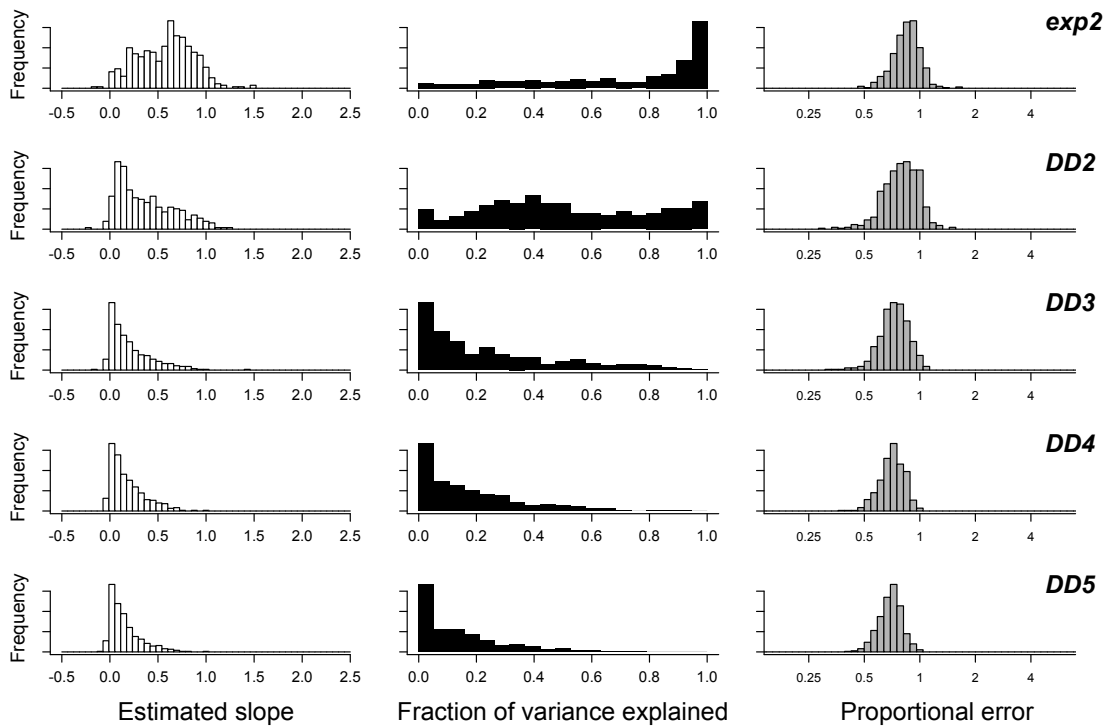

**Figure 8. Precision and bias of MEDUSA in the estimation of branch-specific rates of speciation.** For each simulated phylogeny, MEDUSA was used to estimate the number, location, and parameters of diversification rate shifts. The resulting branch-specific rates

990    of speciation were compared with the true branch rates from the generating model. Results are based on the same simulated datasets analyzed with BAMM and can be directly compared to those shown in Figure 6. Branch-specific speciation rates estimated with MEDUSA show little correspondence with true rates when rates vary through time, at least in comparison to rates estimated with BAMM.

995

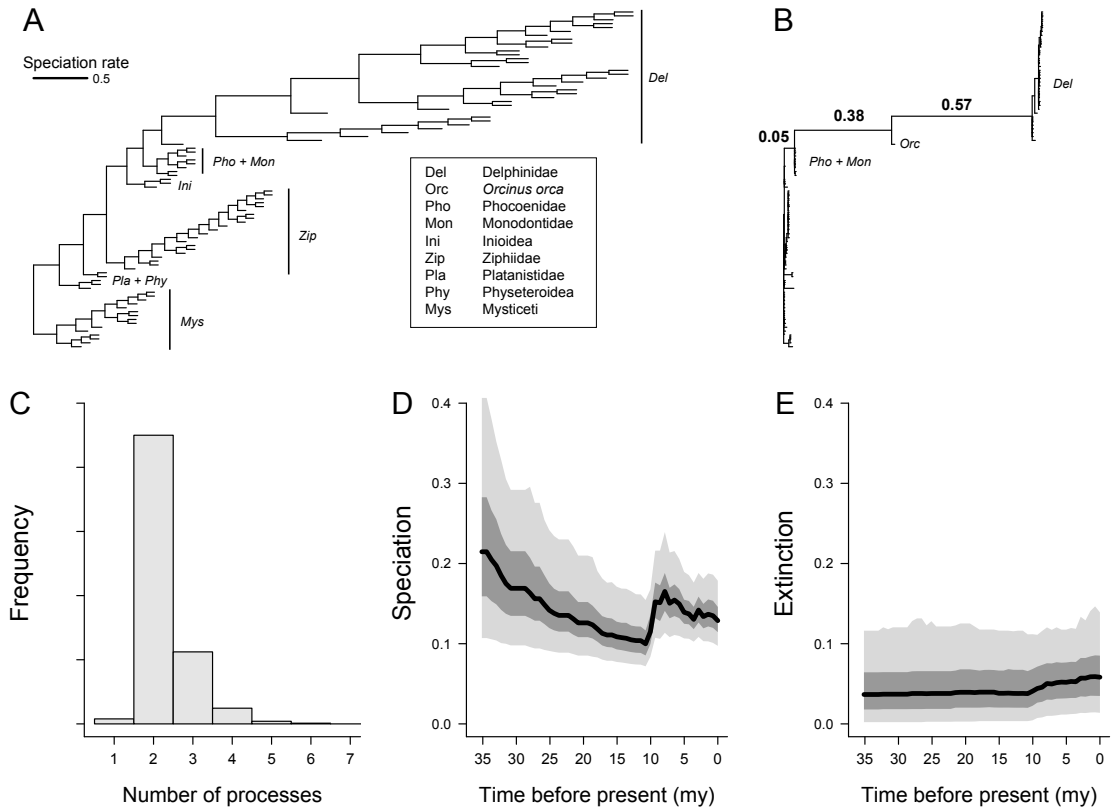

**Figure 9. Dynamics of cetacean diversification through time as revealed by BAMM analysis**. (A) Phylogeny of cetaceans [51], with branch lengths drawn proportional to their marginal speciation rate as estimated using BAMM. A large increase in the rate of speciation (> 6-fold) occurred in one of the ancestral branches leading to the Delphinidae (including or excluding the killer whale, *Orcinus orca*). Despite this increase, the overall trend is towards decelerating rates through time. (B) Cetacean phylogeny with branch lengths scaled by the posterior probability that they contain a rate shift. Numbers above branches denote branch-specific shift probabilities. The probability that a rate shift occurred on at least one of these three branches was 0.975. No other branches had shift probabilities exceeding 0.02. (C) Posterior distribution of the number of distinct processes (including the root process) on the cetacean phylogeny. A two-process model vastly outperforms a one-process model. (D) Speciation rates through time during the extant cetacean radiation; distinct shaded regions denote (from bottom) 0.05, 0.25, 0.50, 0.75, and 0.95 quantiles on the posterior distribution of rates at a given point in time.

Massive spike in mean speciation rates at 7.5 Ma corresponds to the early radiation of the Delphinidae clade. (E) Corresponding extinction through time curve.

1015

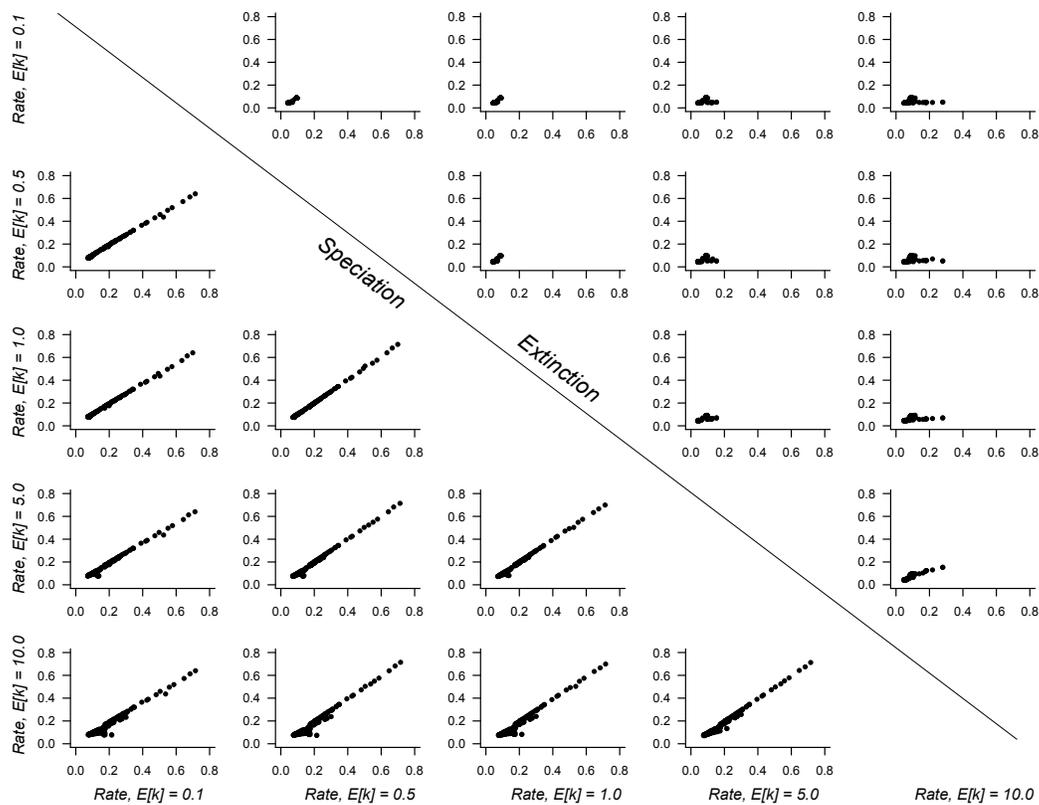

**Figure 10. Sensitivity of marginal rate estimates to prior on Poisson rate parameter.**
Each panel shows a pairwise plot comparing branch-specific (marginal) diversification rate estimates for two values of γ for the Cetacean dataset, with results for speciation and extinction separated by the diagonal. Speciation rate estimates for the cetaceans are remarkably robust to choice of prior: even γ = 10 and γ = 0.1 yield strikingly similar marginal distributions for branch-specific speciation rates. This is generally not true for extinction, where mean marginal rates for each branch were more sensitive to prior formulation. However, extinction was nonetheless estimated to be low overall regardless of the prior γ.

**Table 1. Relationship between branch-specific BAMM estimates of extinction and true rates in the simulation model**.

| *Model* | *processes* | *slope* [a] | *$r^2$* | *PE* [b] |
|---|---|---|---|---|
| Exponential change (exp2) | 2 | 0.76 | 0.59 | 1.27 |
| Diversity-dependent (DD2) | 2 | 0.81 | 0.16 | 1.85 |
| Diversity-dependent (DD3) | 3 | 0.75 | 0.11 | 1.68 |
| Diversity-dependent (DD4) | 4 | 0.81 | 0.13 | 1.67 |
| Diversity-dependent (DD5) | 5 | 0.82 | 0.10 | 1.66 |

[a] Slope and $r^2$ denote the estimated slope and variance explained by the relationship between true and estimated extinction rates for 500 trees simulated under each model.

[b] PE is the mean proportional error across all simulations under a given model.